# Analysis on natural ventilation of the refuge floor in a High Rise Building

Hasarinda Kariyawasam[1], Dileepa Withanage[2], Harith Konara[2], Chathura Ranasinghe[2]

1 Department of Engineering Technology, Sabaragamuwa University of Sri Lanka, P.O. Box 02, Belihuloya, 70140, Sabaragamuwa Province, Sri Lanka.

2 Department of Mechanical Engineering, University of Moratuwa, Bandaranayake Mawatha, Moratuwa, 10400, Western Province, Sri Lanka.

Contributing authors: hasarinda@tech.sab.ac.lk; dileepakasun.dk@gmail.com; harithkavinda0@gmail.com; chathurar@uom.lk;

**Abstract**

The refuge floors have been introduced into high-rise buildings with the aim of ensuring a safe place to stay for the occupants in case of an emergency: primarily during a fire breakout. Provisionals for natural ventilation of the refuge floor have been made mandatory in all fire codes as it is not wise to rely only on mechanical methods to remove the smoke-logged inside the refuge floor. However, the effectiveness of such provisions has been questioned in many studies as it highly depends on factors such as wind direction and building geometry. In this study, using three-dimensional computational fluid dynamics modelling, it is shown that careful analysis of smoke distribution can provide the opportunity to integrate natural ventilation in an effective manner to maintain prolonged habitable conditions on a refugee floor, even without mechanical ventilation. Consequently, the performance of a refugee floor natural ventilation strategy of a proposed 70-storey building is evaluated in this study by evaluating the distribution of temperature, visibility, smoke, and Carbon Monoxide concentration during a fire breakout, with a special emphasis on windows placement and the size of the opening area. Initial windows placement was as per the code requirements and later the windows layout and the opening size were altered step by step based on the simulation findings until a safe configuration is achieved. Results showed that adequate cross-ventilation is a prime factor to establish a safe refugee floor area. Also, fire safety strategies in a building need to be evaluated to confirm their effectiveness since the dominant parameters such as wind direction, building geometry, and building orientation are not identical for each building.

**Keywords— high-rise buildings, refuge floor, natural ventilation, CFD**

**NOMENCLATURE**

Letters and symbols

| | |
|---|---|
| $\boldsymbol{u}$ | : Velocity vector |
| $\omega$ | : Vorticity |
| $H$ | : Stagnation energy per unit mass |
| $\boldsymbol{f}_b$ | : Drag force exerted by the subgrid-scale particles and droplets |
| $\tau$ | : Viscous stress |
| $\tau_{ij}{}^{dev}$ | : Deviatoric part of viscous stress |
| $\delta_{ij}$ | : Kronecker delta |
| $\rho$ | : Density |
| $p$ | : Pressure |
| $K$ | : Kinetic energy |
| $\varepsilon$ | : Rate of viscous dissipation |
| $\mu$ | : Viscosity |
| $S_{ij}$ | : Strain tensor |
| $\mu_t$ | : Turbulent viscosity |
| $\Delta$ | : Filter width |
| $k_{SGS}$ | : Subgrid kinetic energy |
| $\bar{u}$ | : Average value of $u$ at the grid cell center |
| $\hat{u}$ | : Weighted average of $u$ over the adjacent cells |
| $l_{mix}$ | : Mixing length |
| $y^+$ | : Non-dimensional wall-normal distance |
| $u_*$ | : Friction velocity |
| $\kappa$ | : Von Karman constant |
| $z_0$ | : Aerodynamic roughness |
| $L$ | : Obukhov length |
| $\theta$ | : Potential temperature |
| $\theta_0$ | : Ground level potential temperature |
| $\theta_*$ | : Scaling potential temperature |
| $\psi_m$ | : Similarity function of wind model |
| $\psi_h$ | : Similarity function of wind model |
| CFD | : Computational Fluid Dynamics |
| FDS | : Fire Dynamic Simulator |

HVAC : Heating, Ventilation, and Air Conditioning

LES : Large Eddy Simulation

HRR : Heat Release Rate

## INTRODUCTION

High-rise buildings are the most effective solution for utilising limited space in densely populated cities. If the height between the lowest level that a fire vehicle can access and the highest habitable floor is higher than 30 m it is defined as a high-rise building. It is called a super high-rise building if that height is greater than 60m [1]. The most concerned question in high-rise buildings is how to evacuate occupants in case of an emergency. A recent fire broke out at the Bronx apartment in New York, USA in January 2022 causing 17 deaths [2]. Fire at the Kaohsiung building in Taiwan in 2021 killed 64 people [3]. These high-rise fires, in addition to many listed in [4] expose the extent of destruction that could occur and raise concerns about the effectiveness of fire and smoke control and rescue methods in skyscrapers.

**The Refugee Floor**

The "refuge floor" is one such mandatory safety measure in high-rise buildings. The refuge floor is designed to be a safe place for occupants to rest during an evacuation or stay until rescue teams arrive, especially for elders, children, and people with disabilities. In addition, it is a sub-base for firefighting operations and a command point for rescue teams. Also, refuge floors could provide access to an alternative staircase through the refuge floor if one of the proceeding staircases is blocked by smoke, fire, or obstruction and it is an assembly point for occupants in case the escape routes are completely obstructed [5].

As per many standards [1,6,7] design of a refuge floor is governed by a set of common guidelines. A refuge floor should be provided for every 10 floors commencing from the highest habitable floor for a building exceeding 60 m in height. The holding area should be sufficient for 50% of the occupant load of 10 floors above and 0.5 m$^2$ floor area should be reserved for a person. This area should be separated from the rest of the building by using compartment walls having a minimum fire-resistance rating of 2 hours. There should be an external corridor or a smoke-stop lobby to connect the holding area to the building. Emergency lighting should be provided to this floor and that should be connected to a secondary power supply such as a generator or battery etc. Emergency lighting should be energized within 15 seconds if a failure is occurred in the main power supply [1]. Furthermore, the minimum height of refuge floors should be 2.3 m [6].

Further, there are specific guidelines for refugee floor ventilation. Permanent openings should be in the refuge floor to allow natural cross ventilation and that will prevent the smoke from logging inside the floor. Permanent openings are generally located at opposite sides on the refuge floor to induce natural ventilation and the size should be at least 25% of the refuge floor area. The minimum height of the openings should be 1.2 m and those openings should be at least 1.5 m horizontally and 3 m vertically away from any adjoining unprotected opening. And all parts of the holding area should be placed within 9 m of any ventilation opening [1].

In addition, some supplementary recommendations are also provided in local standards. According to the fire code in Hong Kong, the open side of a refuge floor should have at least a direct or diagonal distance of 6 m from an opposite side of a street, the boundary of another site, any other external wall having a less fire-resistance rating, any other building on the same site. Also, the spandrel feature with a height of 0.9 m should be introduced to separate the refuge floor from the below floor to reduce the risk of external fire or smoke spread and there should be a horizontal projection of 0.5 m between these two floors [6]. In Singapore, refuge floors are located every 20 storeys. The area assigned for one person is 0.3 m$^2$. Also, sprinkler systems should be installed if there's any non-residential room on the refuge floor [7].

It should be noted here that, in most fire hazards, people lose their lives due to inhaling toxic gases rather than due to direct contact with fire [8]. Therefore, smoke control within the refugee floor is a paramount design criterion. Wind direction, building architecture, and location of the fire are the main factors deciding the smoke spread pattern [9]. Such information is important for building designers to provide reliable ventilation strategies for any emergency.

Accordingly, the present study analyses the internal spread of fire and smoke in the 376 m tall 70-storey high-rise building, located in Colombo Sri Lanka. Evaluation of the present window arrangement on the refuge floor was assessed and suggested further modification for the window arrangement. Survivable conditions of the refuge floor were enhanced with the aim of increasing the time of evacuation from the building.

**LITERATURE REVIEW**

Practical experiments could be performed to analyse the functionality of fire safety methods used in high-rise buildings. Since these experiments are costly and time-consuming, an alternative solution is required. Although reduced-scale experiments could be used, they produce a limited number of results. Hence numerical simulations are the apparent solution [10]. However, constructing an accurate mathematical model would be a complicated task. Fire Dynamics Simulator (FDS) is a widely used CFD code to model fire situations [10,13,14]. Mathematical models provided in FDS have been validated in many studies [15] for different fire scenarios such as building fires, compartment fires, and tunnel fires.

Ohbaa et al. [16] have experimentally investigated the airflow patterns that generate inside a room due to cross ventilation and derived how airflow bent downward at the front by the front eddy and move upward at the leeward side due to the recirculating eddy at the rear. The results also indicated that the behaviour of wind is 3-dimensional, and it consisted of severe pressure gradients, flow separations, re-attachments, and vortices, then it became highly turbulent with the growth of height [16]. Many studies have revealed that the presence of the refuge floor in the mid-height does not noticeably interfere with the surrounding wind patterns of the building [18,19,20]. Detailed analysis of the effect of the refugee floor openings has been performed by Cheng [19] and results showed that the use of only one opening side could generate severe damage to the ventilation and an increase in the opening size of sidewalls would enhance the ventilation inside the refuge floor. Kindangen et al. [21] have investigated the impact of the roof shape and its orientation on the airflow pattern inside the building. Results showed that velocity magnitude was directly affected by the roof shape and building overhang and roof height also controlled the airflow pattern inside the building.

Several studies have been performed by Cheng et al. [11,12] to analyse the effect of wind direction on the airflow inside the refuge floor. The results indicated that cross-ventilation could be achieved from most wind angles. When the wind blew normally to the building smoke could log inside the refuge floor [12]. Therefore, such wind angles should be omitted, or mechanical ventilation should be introduced [11,18,22]. Cheng et al. [11] investigated the relation of the geometry inside the refuge floor especially the shape and number of service cores to the airflow behaviour inside the refuge floor. It was shown that multiple service cores induce more complications to the airflow than a centered service core. Also, rectangular service cores have been encouraged because circular cores may increase the reverse flow of wind behind the service core at some wind angles [11].

Positions of the service core and columns greatly influence the airflow patterns around the refuge floor [20]. Channel-like flow inside the refuge floor is an advantage to ensure a smoke-free environment because it enhances direct ventilation and induces flow circulation in windward and leeward parts of the refuge floor [23]. The door entrance to the refuge floor would be safer if it is located at sidewalls because the rectangular service core generates channel flow at sidewalls [11,19]. The magnitude and the gradient of the wind flow are influential parameters to induce cross-ventilation inside the refuge floor [9]. The height of the refuge floor should be larger than 2.3m because, at lower heights, the wind flow strength and the pressure gradient between windward and leeward sides are weaker. An increase in refuge floor height could enhance natural ventilation but it would not be feasible from design perspective [22].

Lo et al. [9] have shown that if the origin of the fire is at the floor immediately below the refuge floor, the smoke exerted by the fire floor will migrate to the refuge floor from the openings of the floor with the influence of the wind. Several strategies were implemented to avoid this phenomenon. Lau et al. [13] have studied the use of drencher systems for the openings of the refuge floors as a key solution to avoid entering smoke from the outside of the building. The drencher system creates a water curtain using water heads that are energized by firewater tanks in the building. However, Chow et al. [24] have suggested drencher systems are not necessary since the existing methods can avoid smoke logging even for large fire loads [6]. In addition, the spandrel at the window opening could act as a fire-rated partition since it could limit the movement of fire into the refuge floor [3]. The propagation of fire along the external wall has been analysed by Satoh et al. [25]. The results showed that the oscillatory motion of fire is governed by the heat release rate of the fire, and it is weekly affected by window configurations such as window height, soffit height, and balcony height. Sugawa et al. [26] derived empirical equations for upward velocity, temperature, and flame angle referring the experimental results of the ejected fire plume from the external side wind. Soltanzadeh et al. [27] have evaluated the performance of the refuge floor with other evacuation methods. Results showed that refuge floors increase the evacuation rate and reduce the queue of elevator evacuees. Fire codes especially in the USA, UK, and China specify elevators as an occupant evacuation method because the increase in height of the building would increase the evacuation time if the stairs are the only method of evacuation [28]. Cai et al. [28] have suggested a multiple-level pressurization system considering the fire floor to be more effective than regular smoke extraction and staircase pressurization. Occupant behaviour was analysed in studies using eye-tracking in experimental evacuation and Virtual Reality (VR) experiments [28,29]. Results of the studies by. Mossberg et al. [28] have revealed that exit signs pointed to the fire lifts could be an influential factor to promote elevator evacuation. Ding et al. [30] studied the combination of stairs and elevators for evacuation and revealed that there is an ideal percentage of evacuation by elevators that was not related to the number of occupants removed from the elevators.

Although there are many studies analysing the airflow patterns on the refuge floor, there is only very limited research available on fire and smoke movement on the refuge floor. In fact, most studies have limited to scaled models, and analysis of full-scale actual buildings is very rare [16-20]. Furthermore, no work has been reported on the effect of window arrangement on smoke and fire spread on a refugee floor. Consequently, the present work aims to address these gaps by analysing the effect of window arrangement in a refugee floor on the smoke spread of a proposed 70-storey commercial building.

## NUMERICAL APPROACH

For the work presented here the FDS simulation software was used. FDS solves the governing equations in a uniform rectilinear grid. Large Eddy Simulations (LES) were carried out with the reaction progress variable approach for combustion calculations. Eddy Dissipation Concept (EDC) was selected to simulate the reaction rate and infinite fast chemistry was assumed. Radiation heat transfer effects were also considered for calculations.

The momentum equation solved in three dimensions is given by,

$$\frac{\partial \boldsymbol{u}}{\partial t} - \boldsymbol{u} \times \omega + \nabla H - \tilde{p}\nabla\left(\frac{1}{\rho}\right) = \frac{1}{\rho}[(\rho - \rho_0)\mathrm{g} + \boldsymbol{f}_b + \nabla.\tau] \qquad (1)$$

Where $(\boldsymbol{u}.\nabla)\boldsymbol{u} = \nabla|\boldsymbol{u}|^2/2 - \boldsymbol{u} \times \omega$ and $H \equiv |\boldsymbol{u}|^2/2 + \tilde{p}/\rho$

Here, $f_b$- drag force exerted by the subgrid-scale particles and droplets, $\tau$- viscous stress

The transport equation of the resolved kinetic energy is given by,

$$\bar{\rho}\frac{DK}{Dt} + \frac{\partial}{\partial x_j}([\bar{p}\delta_{ij} + \tau_{ij}^{dev}]\widetilde{u}_i) = \bar{p}\frac{\partial \widetilde{u}_i}{\partial x_i} + \tau_{ij}^{dev}\frac{\partial \widetilde{u}_i}{\partial x_j} + (\bar{\rho}\mathrm{g}_i + \overline{f_{b,i}})\widetilde{u}_i \qquad (2)$$

Here, $\tau_{ij}^{dev}$- deviatoric part of viscous stress, $\delta_{ij}$- Kronecker delta

The LES sub-grid closures assume production of sub-grid kinetic energy is equal to the dissipation of total kinetic energy which leads to the equation for $\varepsilon$,

$$\varepsilon \equiv -2\mu\left(S_{ij}S_{ij} - \frac{1}{3}(\nabla.\boldsymbol{u})^2\right) \qquad (3)$$

Here, $\mu$-viscosity, $S_{ij}$- strain tensor

The Deardorff model was used as the SGS turbulence model. The following equations were solved to obtain turbulent viscosity ($\mu_t$).

$$\mu_t = \rho C_v \Delta \sqrt{k_{SGS}} \qquad (4)$$

$$k_{SGS} = \frac{1}{2}\left[(\bar{u} - \hat{u})^2 + \left((\bar{v} - \hat{v})\right)^2 + \left((\bar{w} - \hat{w})\right)^2\right] \qquad (5)$$

Here, $\rho$- air density, $C_v$- model constant, $\Delta$- filter width, $k_{SGS}$- subgrid kinetic energy, $\bar{u}$- average value of $u$ at the grid cell center, $\hat{u}$- weighted average of $u$ over the adjacent cells. $C_v$ value is taken as 0.1.

Wall function was provided to achieve accurate results for velocity near walls. Mixing length ($l_{mix}$), and non-dimensional wall-normal distance ($y^+$) was combined in the equation as follows.

$$l_{mix} = C_s \Delta [1 - e^{-y^+/A}] \qquad (6)$$

Here $C_s$ and $A$ represents model constant and dimensionless empirical constant respectively. Values for the constants in the model are $C_s = 0.2$ and $A = 26$.

The oncoming wind needs to be accurately represented for better results. Monin-Obukhov similarity method [31] was used to model the wind profile suitable for a city area.

$$u(z) = \frac{u_*}{\kappa}\left[\ln\left(\frac{z}{z_0}\right) - \psi_m\left(\frac{z}{L}\right)\right] \qquad (7)$$

$$\theta(z) = \theta_0 + \frac{\theta_*}{\kappa}\left[\ln\left(\frac{z}{z_0}\right) - \psi_h\left(\frac{z}{L}\right)\right] \qquad (8)$$

Where, $u$ - wind speed profile, $u_*$ - friction velocity, $\kappa$ - Von Karman constant, $z_0$ – aerodynamic roughness, $L$ - Obukhov length, $\theta$ - potential temperature, $\theta_0$ - ground level potential temperature, $\theta_*$ - scaling potential temperature, $\psi_m$ and $\psi_h$ - similarity functions. Model constants are taken as $L = -100$, $z_0 = 2$ and $\kappa = 0.41$.

## VALIDATION OF THE COMPUTATIONAL MODEL FOR A FIRE IN A 10-STOREY BUILDING

The experimental fire reported by Hadjisophocleous et al. [33] in a 10-storey building located in the fire research laboratory of the National Research Council of Canada was used as the case for validation. This building was set up in a manner that the generated smoke is moving through the stair shaft of the building and spreads to the other floors from the service core of the building.

Open/ Closed conditions of the doors in the building are shown in table 1. This door arrangement ensures that fire originates on the 2nd floor and would feed continuously with fresh air. Propane was used as the fuel of the combustion and the fire compartment was located at a compartment that has a volume of (9 x 3.8 x 3.3) m$^3$, and the fire tray has a cross-section area of (5 x 0.4) m$^2$ which is at the center of the fire compartment. Variation of the heat release rate during the fire was measured, and it was included in the mathematical model as a user-defined function (figure 1).

| Table 1: Open/Partly Open/Closed conditions of doors on each floor [32] | | | |
|---|---|---|---|
| Floor No | DR4 | DR5 | Notes |
| 10 | Open | Partly Open | |
| 9 | Closed | Open | |
| 8 | Open | Partly Open | |
| 7 | Closed | Open | |
| 6 | Open | Partly Open | |
| 5 | Closed | Open | |
| 4 | Open | Partly Open | |
| 3 | Closed | Open | |
| 2 | Open | Open | |
| 1 (Ground floor) | Open | Open | DR2 Open DR6 Closed |

Thermodynamics properties of the propane fuel were added to the FDS model as shown in table 2. Also, the same properties of other materials such as concrete, wood, steel, and ceramic fiber coating were included as mentioned in table 3. The temperature at different locations and gas concentration in front of the stairs were selected as the parameters which were measured from the experimental setup and validated using FDS. The locations of thermocouples and gas analysers are shown in figure 2. The domain of this model is (15 x 9 x 28.8) m$^3$, and it is divided nearly into 4 million elements. Therefore, the volume of one cell is (0.1 x 0.1 x 0.1) m$^3$. More information related to this experiment could be found in [34].

| Table 2: Combustion properties of Propane [34] |
|---|

| Fuel | Heat of Combustion (kJ/kg/K) | Soot yield factor $Y_s$ (g/g) | CO yield factor $Y_{CO}$ (g/g) |
|---|---|---|---|
| Propane | 4.37 | 0.024 | 0.005 |

| Table 3: Material properties of the building [34] | | | | | |
|---|---|---|---|---|---|
| Item | Material in FDS | Thickness (m) | Density (kg/m$^3$) | Specific heat (kJ/kg/K) | Thermal Conductivity (W/m/K) |
| Wall/Floor/Ceiling | Concrete | 0.2 | 2200 | 0.88 | 1.37 |
| Stairs | Steel | 0.08 | 7830 | 0.47 | 43 |
| Doors | Wood | 0.1 | 420 | 2.70 | 0.11 |
| Isolation of fire Compartment | Ceramic Fibre | 0.025 | 96 | 1.025 | 0.15 |

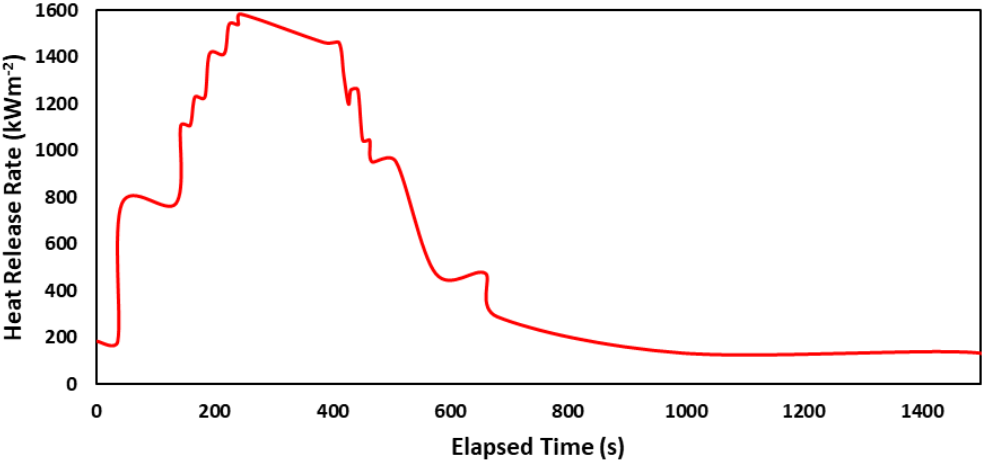

Figure 1: Fire Ramp (HRR vs Time)

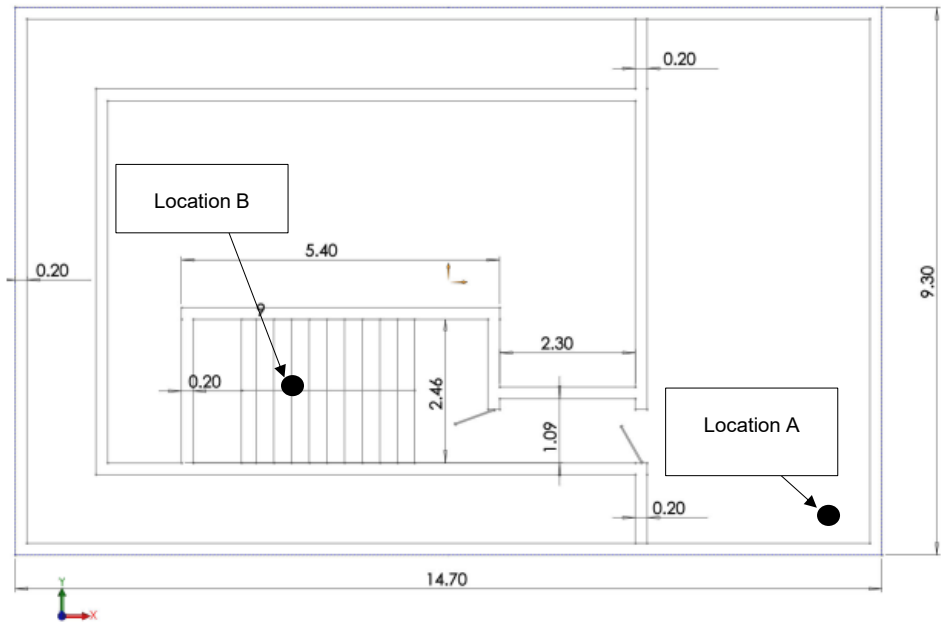

Figure 2: Layout of thermocouple probes and gas analysers inside the building (measured in millimetres)

**Comparison of Results**

Figure 3 shows the comparison of temperature between experimental and simulation results on the fire floor. Figure 3 (a) shows the results at 0.62 m height and 3 (b) shows at 2.57 m height at location A. The temperature has risen with time proportional to the fire intensity. From the comparison of graphs, it can also be seen that the temperature value at increased heights from the floor level is higher due to the accumulation of hot burnt gases. The prediction of these trends by the computational model is quite satisfactory, though some deviation of the absolute value is seen.

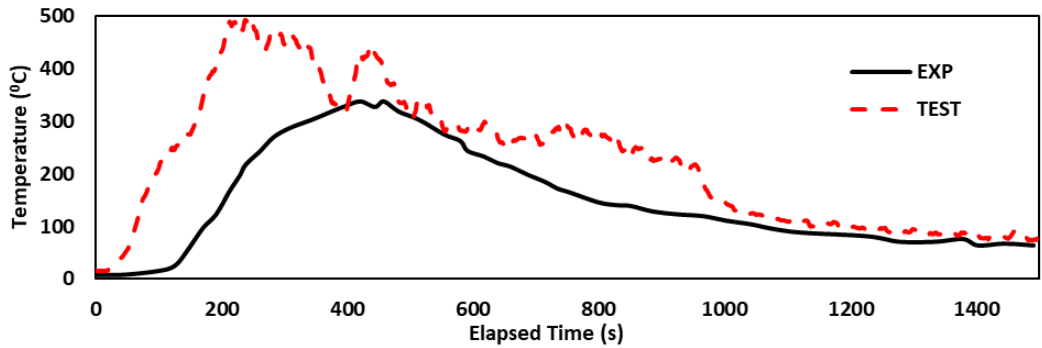

(a)

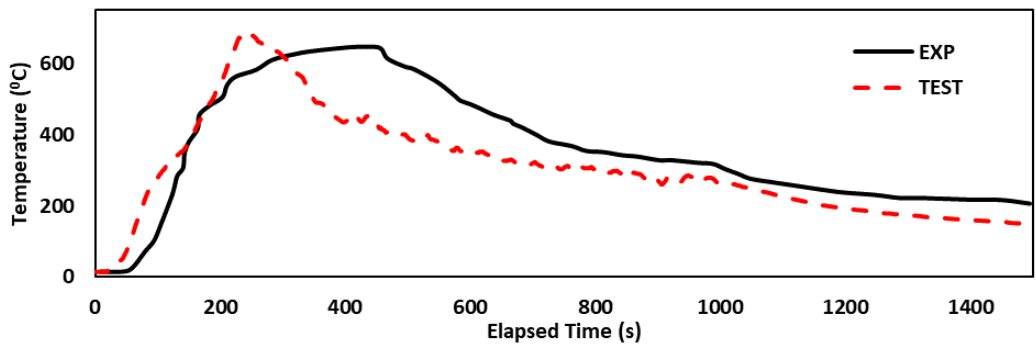

(b)

Figure 3: Variation of temperature with time on the fire floor (2nd floor) at southeast corner; (a) readings at 0.62 m height (b) readings at 2.57 m height

Figure 4 shows the results in the stair shaft on the second floor and fifth floors at location B. It's clear that in figure 4 (a) the simulation couldn't get the high spike at the beginning. However, the results match each other at the end of the graph. In figure 4 (b), both results could be considered in good agreement. Although, in the end, simulation results deviate from the experimental results. There are several possible reasons for this deviation in results. In the experimental building, the fire compartment was covered using ceramic fiber and it was not considered in this model. Also, square-shaped openings that create paths to the surrounding have dimensions of 0.24 x 0.24 $m^2$ [34] but in this model, we could create 0.2 x 0.2 $m^2$ due to the low resolution of the domain and it couldn't resolve due to the lack of computational power. This domain already contains 3.8 million mesh elements.

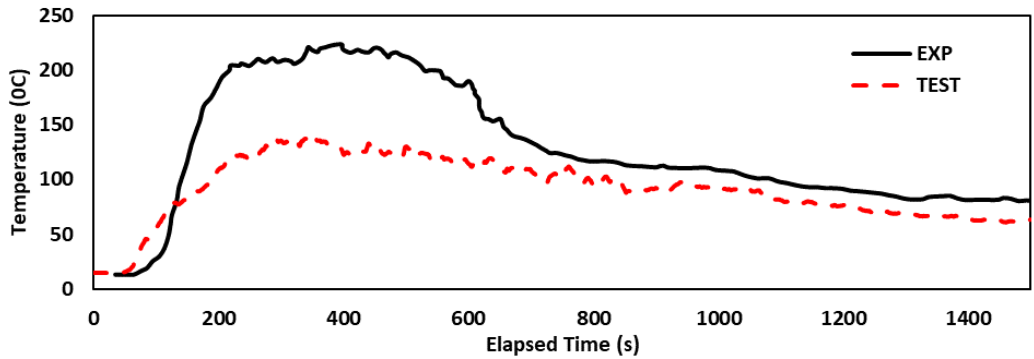

(a)

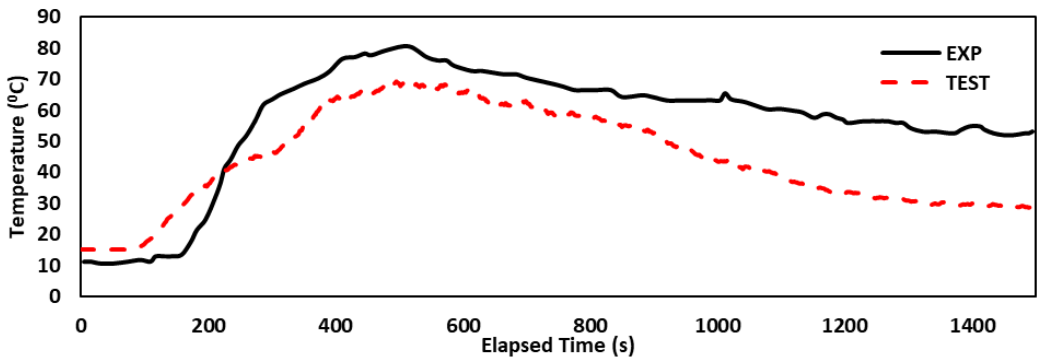

(b)

Figure 4: Variation of temperature with time at the middle of the stair shaft 06 m below the ceiling (a) readings on Second Floor (b) readings on Fifth Floor

**Comparison of Gas concentration**

Oxygen and Carbon Dioxide gas concentration at the stair shaft on the fire floor is shown in figure 5. It could be seen that the $CO_2$ and $O_2$ gas concentrations in the fire floor recovered 1000 seconds after the initiation of the fire. However, the results of this simulation indicate that it isn't fully recovered. The difference in the cross-section areas of the openings on the 4th floor and 8th floor that are mentioned above might be the reason for the deviation of the simulation results.

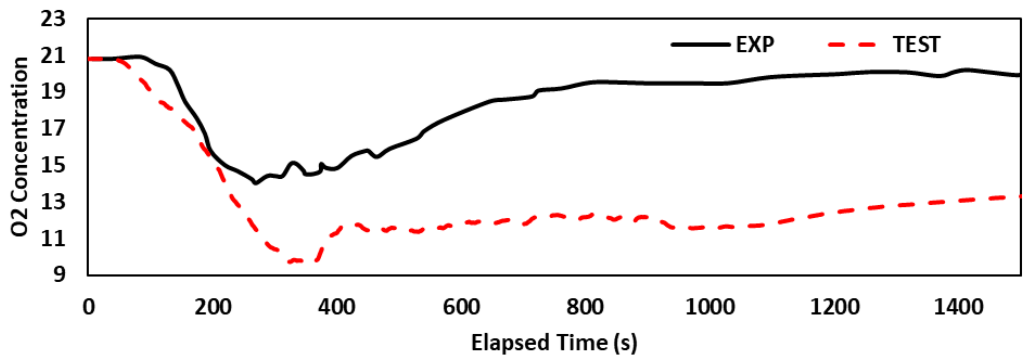

(a)

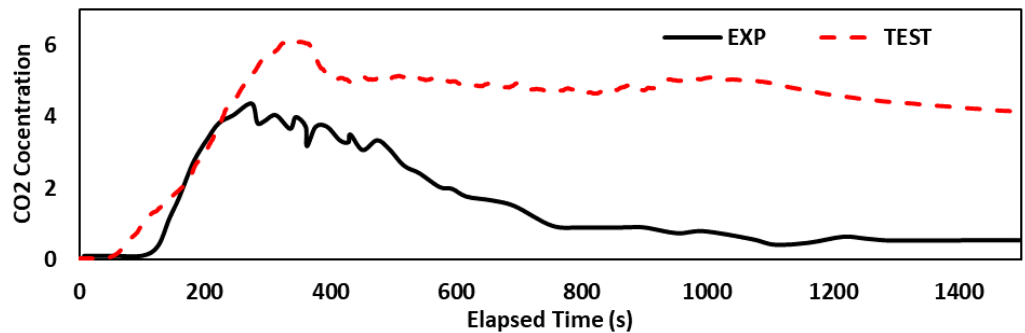

(b)

Figure 5: Variation of gas concentrations with time in the stair shaft at the fire floor
(a) $O_2$ gas concentration (b) $CO_2$ gas concentration

## SMOKE SPREAD IN THE REFUGEE FLOOR

### THE BUILDING GEOMETRY AND WIND CONDITIONS

The selected building for the study has a height of 376 m and it would be one of the tallest buildings in South Asia. This mixed occupancy building facilitates hotels, offices, and residential and retail facilities. Two refuge floors are on 33rd and 52nd floors of this building and the smoke distribution on the 33rd floor was analysed in this study. It is expected that a population of 1149 occupants would be accommodated between the 33rd floor and 51st floor. Then, this refuge floor should provide room for 50% of the occupants by allowing space of 0.5m$^2$ per person. The building design reserves an area of 666.03 m$^2$ for the refuge floor which is well above the requirement (287.5 m$^2$). A building height, 10 floors above the 33$^{rd}$ flow refugee area from the ground level were considered as the problem domain (figure 6). The remaining floors were not considered for the simulation domain as they are too far from the interested area (33$^{rd}$ floor) to induce a noticeable effect on the flow inside the refuge area.

Annual weather reports of several years were referred to assign 30 ℃ for atmospheric temperature and 4.5 ms$^{-1}$ for wind velocity at a height of 10 m [35]. Concentrations of Soot and CO were assigned as 0.011 and 0.038 respectively based on the [36]. Experimentally estimated fire ramp by Rinne et al. [32] was chosen to model the heat and exhaust gas emission rates at the fire location.

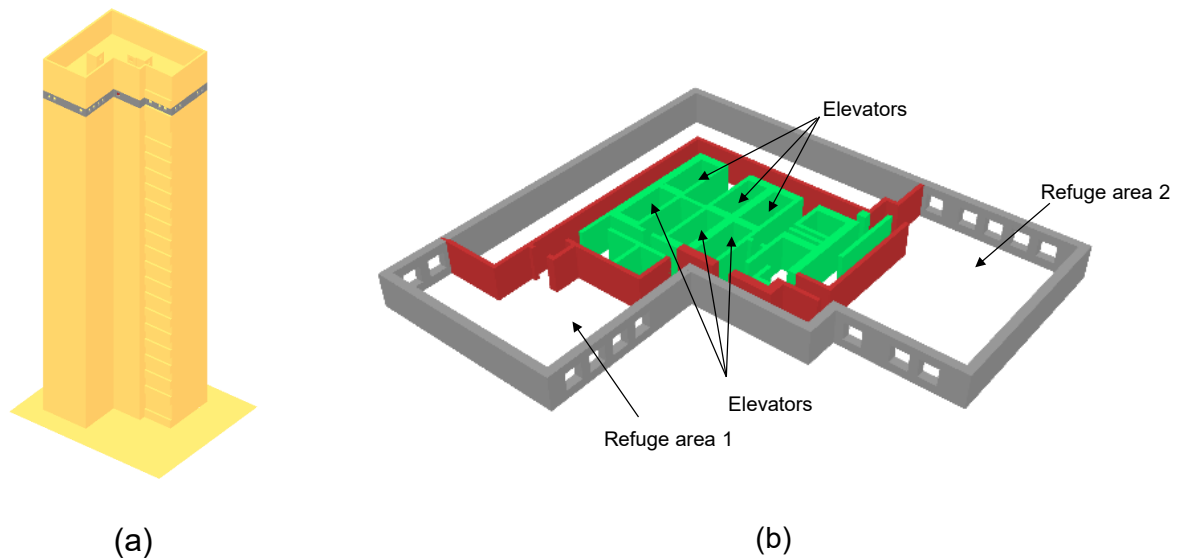

(a)  (b)

Figure 6: The selected building: (a)Location of the refugee floor is marked in ash colur (b) refuge floor layout

Based on wind data, wind-blowing directions could be narrowed down to the West and North-East [35]. The fire was assumed to be originated in the staircase at the floor below the refuge floor. The fire propagation path was from the staircase through the service core at the refuge floor to the refuge space. The worst-case scenario was considered for simulations. It was assumed that the electric system had failed therefore the pressurization of the staircase was stopped. All fire doors were assumed to be failed at the beginning and these doors acted as openings.

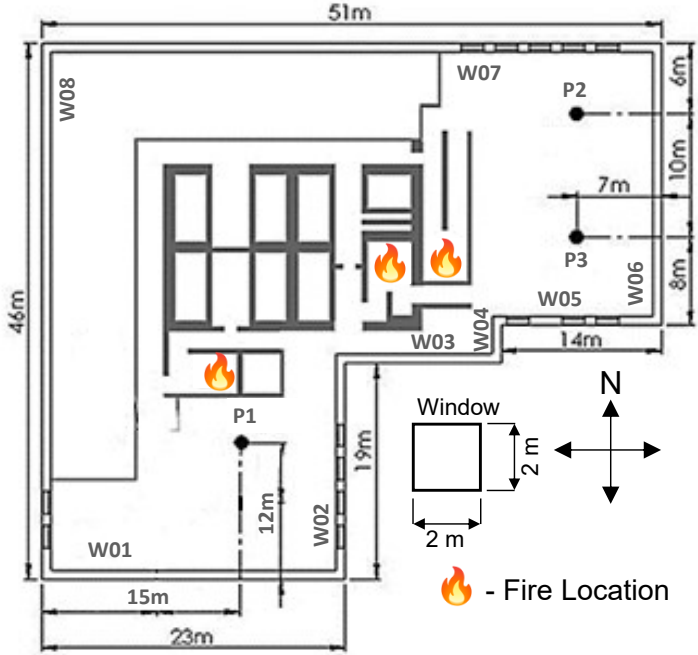

Figure 7: Measuring points of the Refuge Floor – Window configuration - 0

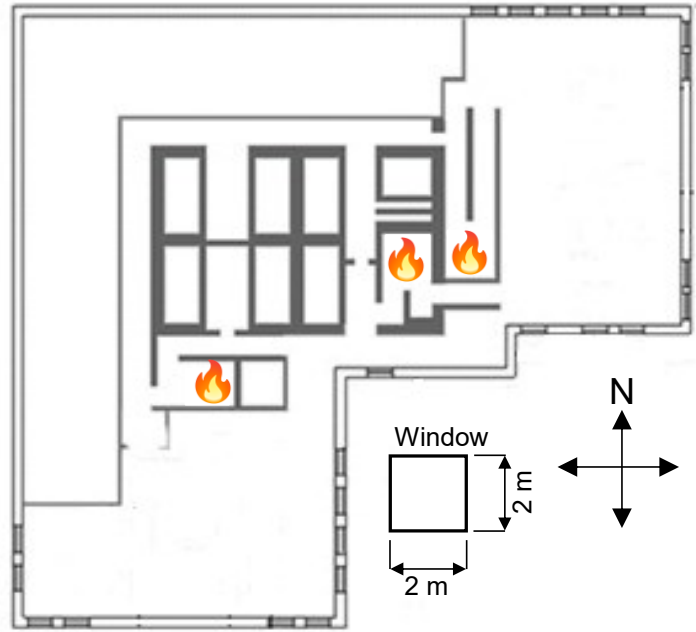

Figure 8: Window configuration - 01 of the refuge floor

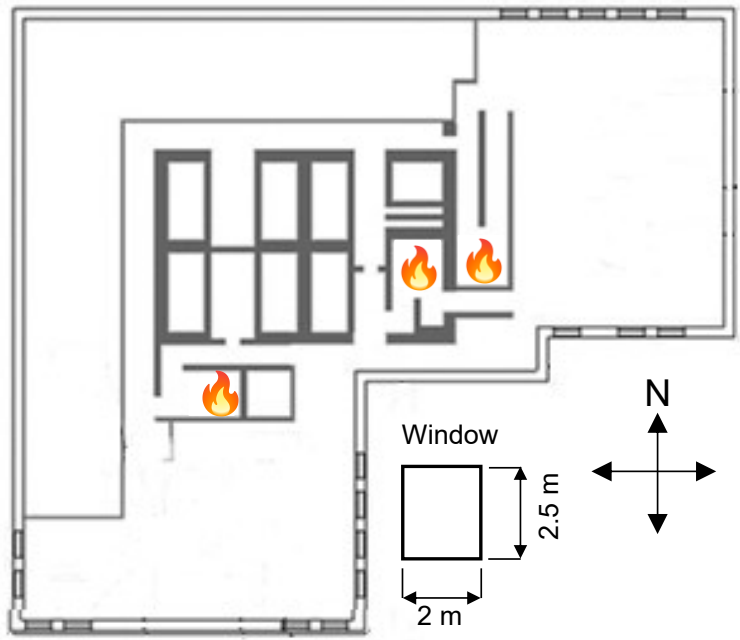

Figure 9: Window configuration – 02 of the refuge floor

The proposed window configuration by the building designers (configuration - 0) was evaluated at first and contrasted with two other modified window configurations (1 and 2) that could enhance the occupant's safety inside the refuge floor. As mentioned above, the area of permanent openings is 13% of the floor area in the window configuration - 0 which was suggested by the building designer. Configurations 01 and 02 are proposed for the analysis by increasing the ratio up to 25%. Both configurations possess more windows than configuration - 0 while the height of the windows in configuration - 2 is higher (2.5 m) than in the other two setups. Window configurations are shown in figures 7-9.

Smoke layer concentration, Carbon Monoxide (CO) concentration, temperature, airflow velocity, and visibility were analysed during the fire and smoke spread. The refuge floor was divided into two sections for the investigation as shown in figure 6. Three locations were selected for the analysis as shown in figure 7 considering the geometry of the floor (refuge area 1 and 2) and the criticality of the air quality.

Modifications were proposed by analysing the following weaknesses in the current window configuration. There was an accumulation of smoke in refuge area 1 in the designed window setup. Configuration - 1 was able to keep smoke logging at a minimum in refuge area 1 and improved the survivable conditions. The concept of configuration - 1 was based on the results of the designed setup. The number of windows was increased as it was necessary to comply with the newest fire regulations. An area of windows was added considering the increment of cross ventilation. Then, extreme velocities in refuge area 1 were detected in the current window setup. So, three windows were introduced to wall 01 to reduce the wind velocity. Also, there was a vortex created at refuge area 2 when the wind blew from the West direction. So, windows were introduced at wall 06 for configuration - 1. However, this vortex kept occurring in configuration - 1 as well. Therefore, the windows at wall 06 were removed and that opening area was added to the existing windows in configuration - 2.

## RESULTS AND DISCUSSION

### Discussion using data on temperature

Temperature is the prominent parameter to discuss the safety of the refuge floor. Figure 10-(a) depicts the temperature distribution at 1 m height on the refuge floor when the wind was blowing from the North-East direction, and it was obtained 600 seconds after the ignition. There were three distributions in figure 10 (a)(i)-(iii) which give information about the three window configurations when the wind blew from the North-East direction. It is clear that the modifications (1 and 2) act better than the designed configuration (0) especially in refuge area 1. Figure 10 (b) represents the same results for the West direction. Overall, configuration 2 acted better than others. However, the temperature at refuge area 2 was way over the limit (40 ℃) [38] which meant refuge floor area 2 was not survivable when the wind was from the West direction.

Since figure 11 indicates the temperature at 3 m height, it is clear that the temperature has risen with the height. From figure 11 (a) in which the wind blew from the North-East, configuration 1 acted better than the other. The major difference in configuration 1 was that it had four windows on wall 06 while there were not any in others. However, configuration 2 acted better when the wind was from the West due to the same difference.

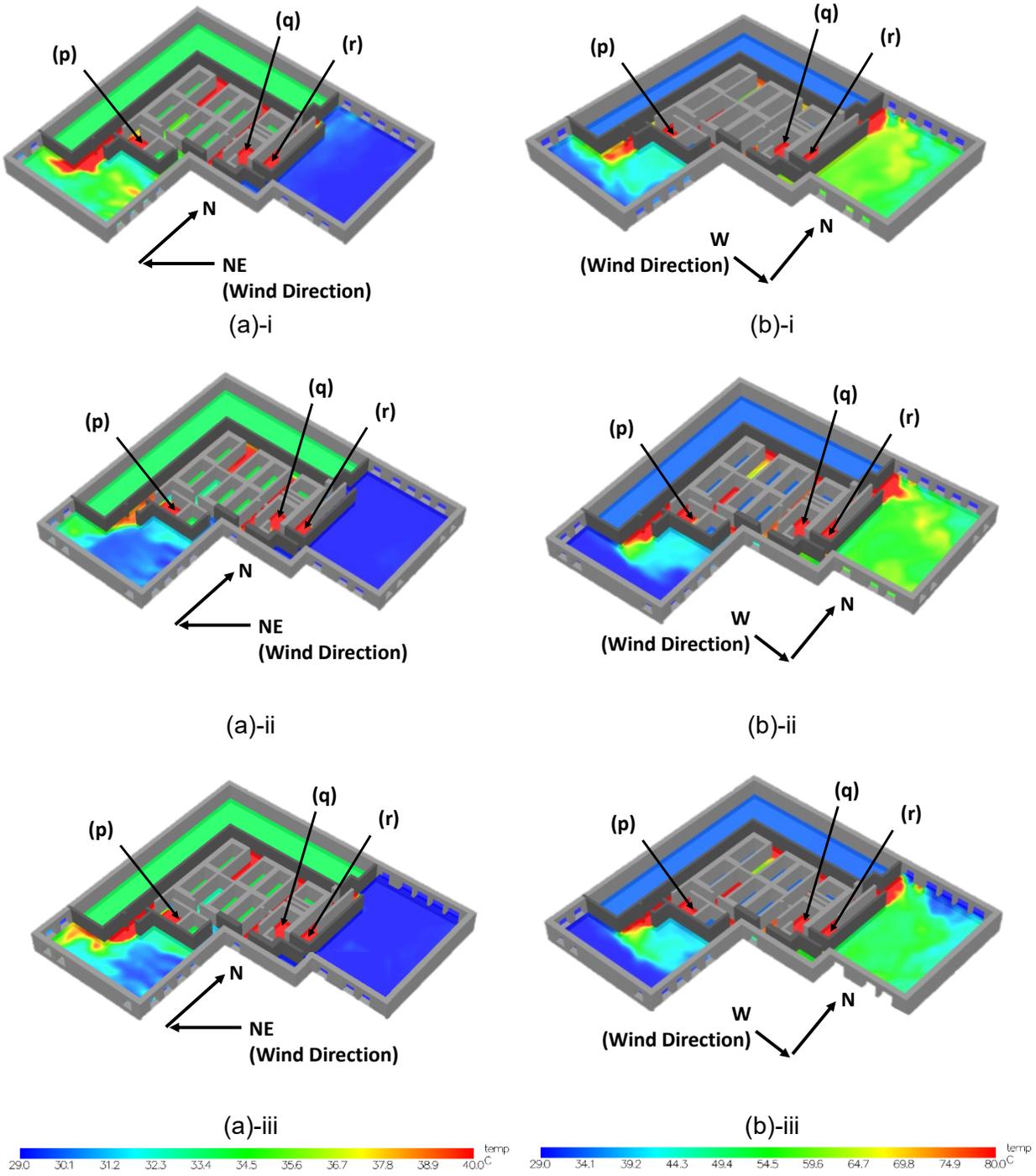

Figure 10: Temperature distribution on the refuge floor at 1 m height after 600 seconds of fire; (a)- wind is from North-East (b)- wind is from West; i-current setup, ii-1st modification, iii-2nd modification; p, q, r are the fire locations

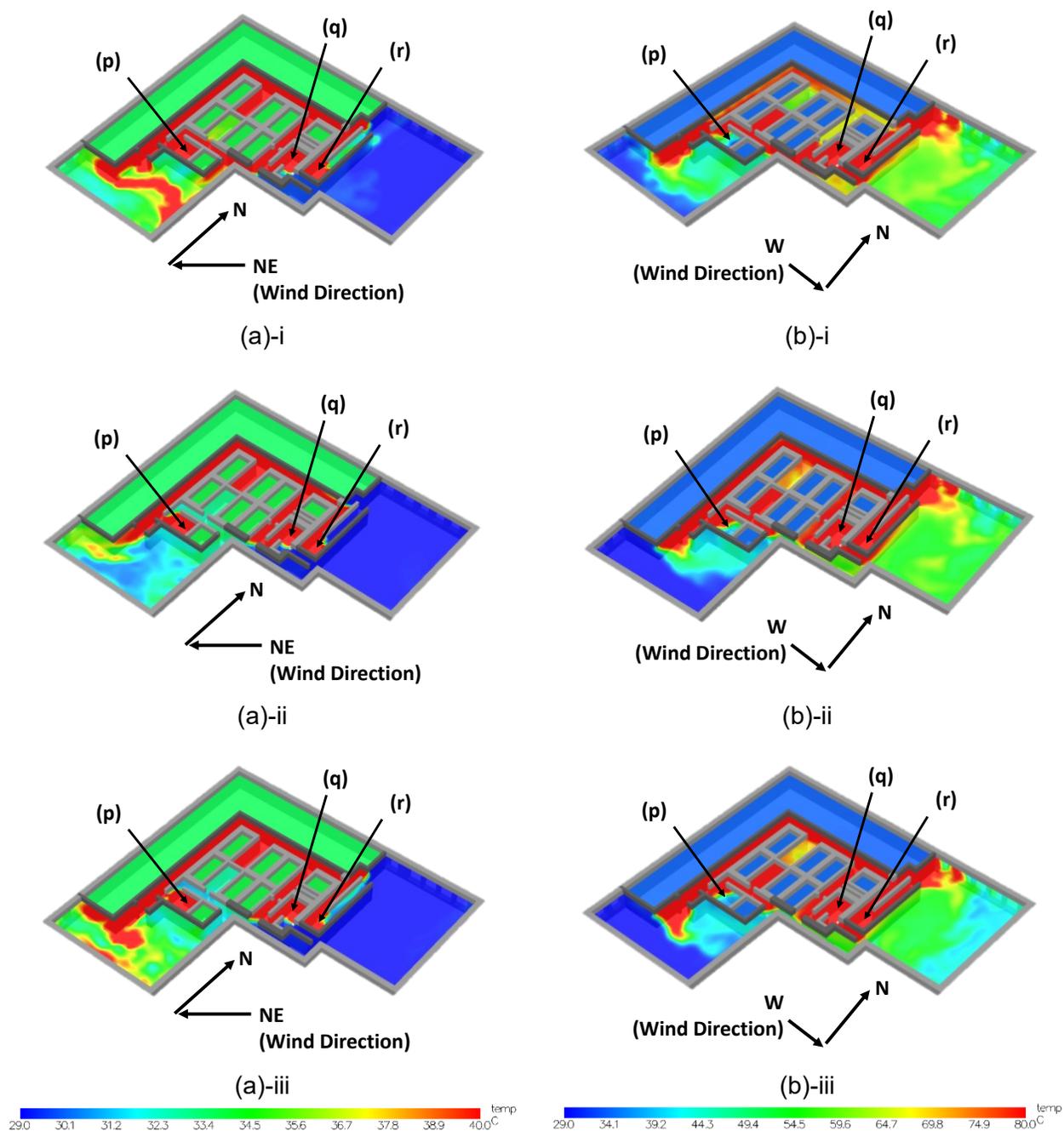

Figure 11: Temperature distribution on the refuge floor at 3 m height after 600 seconds of fire; (a)- wind is from North-East (b)- wind is from West; i-current setup, ii-1st modification, iii-2nd modification; p, q, r are the fire locations

Figures 12 and 13 show a graphical representation of 20 seconds moving average temperature vs time considering the window arrangement and height. Overall, it could be seen that modifications worked better than the designed window setup. The designed window configuration failed to keep under 40 °C in figure 12 when the wind was from the North-East while configuration - 1 showed the best performance. Figure 13 represents the variation of temperature with height, and it was an interesting observation that the minimum temperature was at 2 m height. The maximum velocity of wind inside the refuge floor was at 2 m height because the windows were positioned from 1 m to 3.5 m height. Therefore, the impact from the cross-ventilation was maximized in the middle (2 m). This behaviour was beneficial to stop the logging of smoke at the ceiling since the rising smoke would be caught by the moving wind.

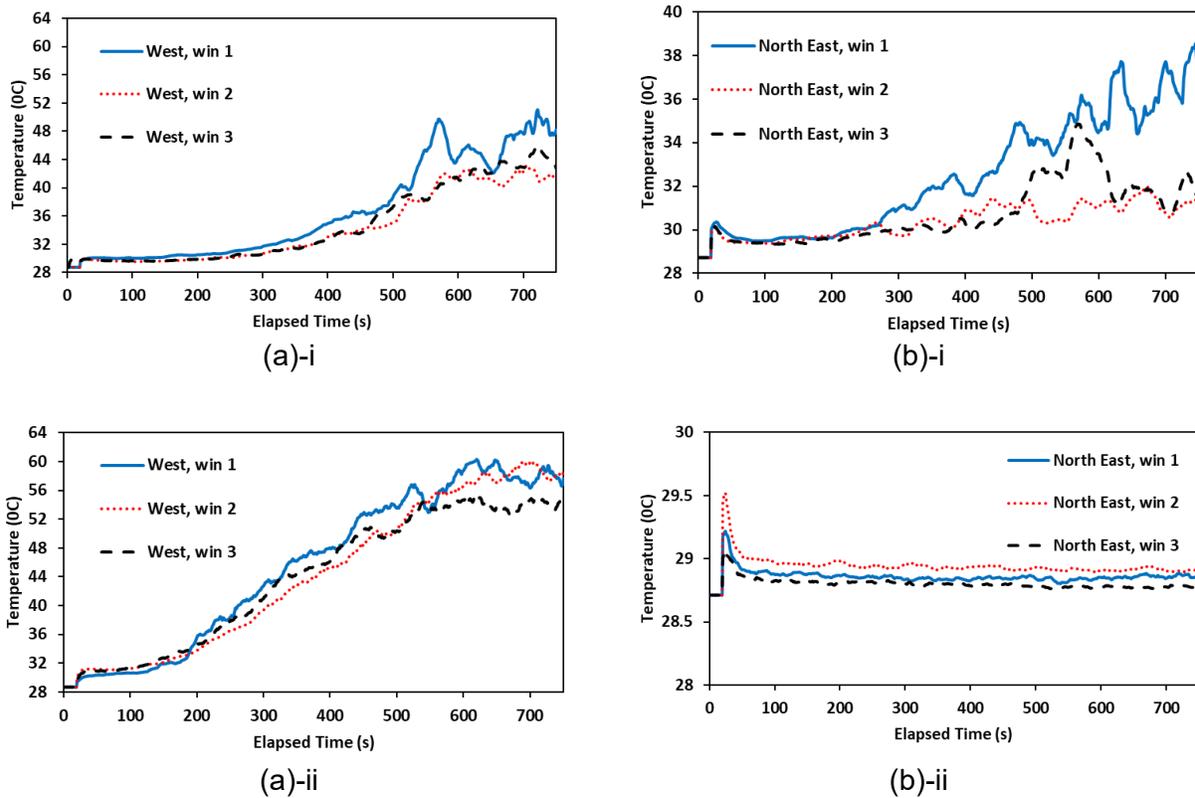

Figure 12: Variation of temperature with time of different window setups at 2m height; (a) readings from P2 in refuge area 1 (b) readings from P3 in refuge area 2; i-wind is from West, ii- wind is from North- East

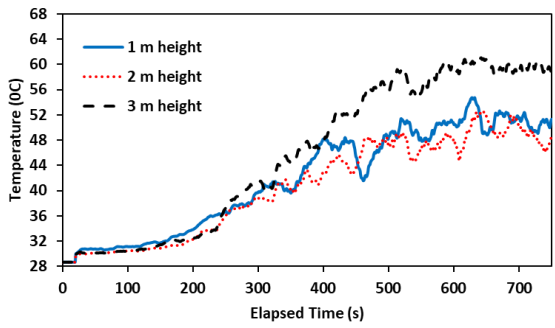
(a)-i

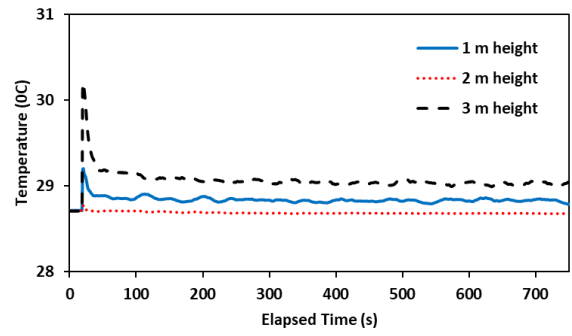
(b)-i

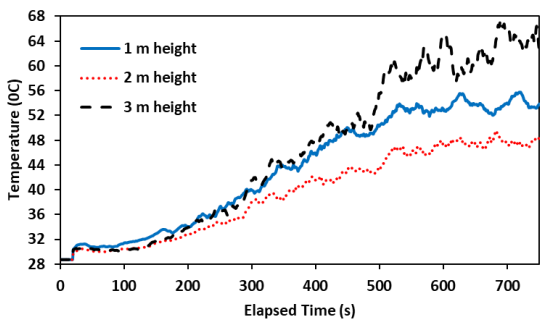
(a)-ii

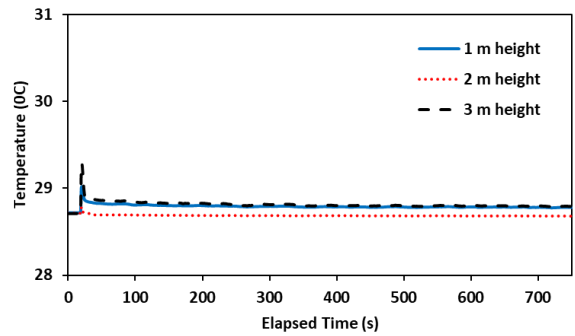
(b)-ii

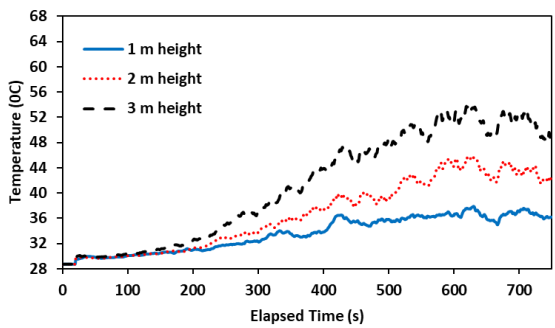
(a)-iii

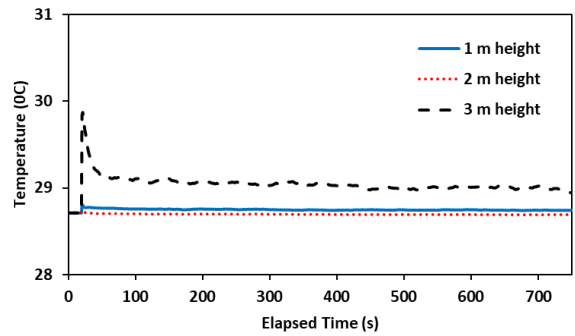
(b)-iii

Figure 13: Variation of temperature with time at different heights on P1; (a) wind is from West (b) wind is from North-East area 2; i- current setup, ii- 1st modification, iii- 2nd modification

**Discussion using data on concentration of Carbon Monoxide**

CO is a toxic gas and the concentration needed to be kept below 9 ppm according to the ASHRAE standards [38]. The results of 20 seconds moving average CO monoxide concentration with time were plotted with time in figure 14. Overall, it is clear that modifications acted better than the designed window configuration. CO concentration was well below 9 ppm in both refuge areas 1 and 2 when the wind is from North-East. All window configurations failed to keep the survivable condition in refuge area 2 when the wind is from the West. It could be concluded that window configuration - 2 showed better results than others.

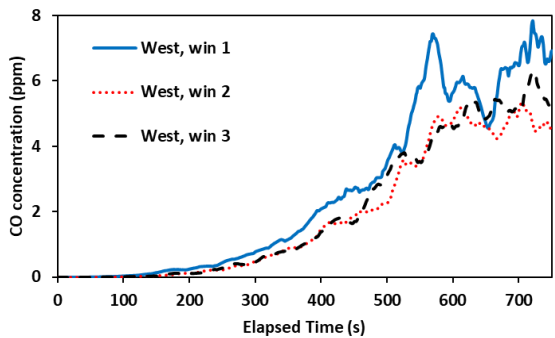
(a)-i

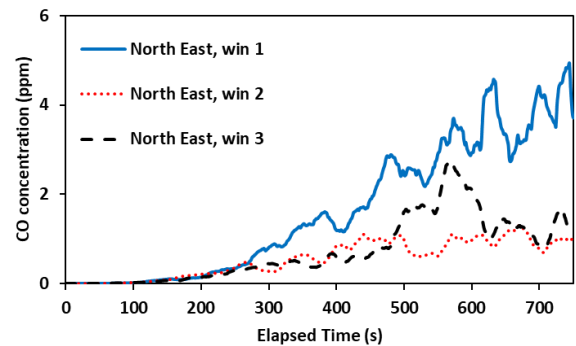
(b)-i

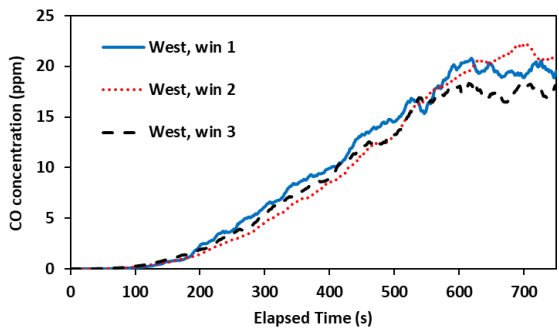
(a)-ii

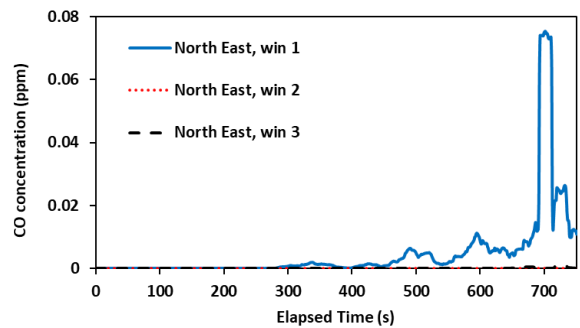
(b)-ii

Figure 14: Variation of CO concentration with time of different window setups at 2m height; (a) readings from P2 in refuge area 1 (b) readings from P3 in refuge area 2; i-wind is from West, ii-wind is from North-East

**Discussion using data on smoke and visibility**

Smoke spread throughout the refuge floor after 600 s of fire is shown in figure 15. Smoke entered the refuge area from the service core. Figure 15 (a) represents the smoke movement when the wind was from North-East. There was no smoke accumulation in refuge area 2 in all window setups. However, refuge area 1 suffered from smoke logging, and configuration - 1 showed the best results compared to others. Windows on wall 06 in configuration - 1 might increase the air volume inside the floor. When the wind blew from the West, smoke densely accumulated in refuge area 2. Also, refuge area 1 was affected by smoke. However, dense smoke was low in configuration - 1 than in the others. Cross-ventilation is the key factor to extract smoke from the refuge floor. When the wind blew from North-East, the window setups in every configuration were capable to induce cross ventilation. However, when the wind is from the West, the window arrangement failed to induce cross ventilation and that led to the logging of smoke inside the refuge floor.

Visibility was derived considering three main parameters namely Soot yield, mass extinction coefficient ($K_m$), and visibility factor. Soot yield is the fraction of fuel mass that is converted to soot where the simple chemistry approach is being used. $K_m$ depends on various light-absorbing gas species. The value of $K_m$ is about 8700 $m^2$/kg for flaming combustion. The visibility factor is a non-dimensional constant (C), and its default value is 3. C varies with the type of object that could see through the smoke. For the light-emitting sign, C = 8, and for the light-reflecting sign, C = 3. The maximum value for visibility is 30 m [31].

Figure 16 shows the variation of 20 seconds moving average visibility vs time. Overall, it is clear that the refuge floor would experience more visibility when the wind blew from North-East. Refuge area 2 was perfectly visible in every window configuration. While the designed window configuration showed some fluctuation in the visibility in refuge area 1, configuration – 1 could keep the visibility at the maximum during the considered time. Configuration – 1 could have more visibility than the other two configurations at around 20 m in refuge area 1 when the wind was from the West. Although the visibility in refuge area 2 is gradually decreased in each setup with the increase of heat release rate and reaches 5 m when the fire is fully developed.

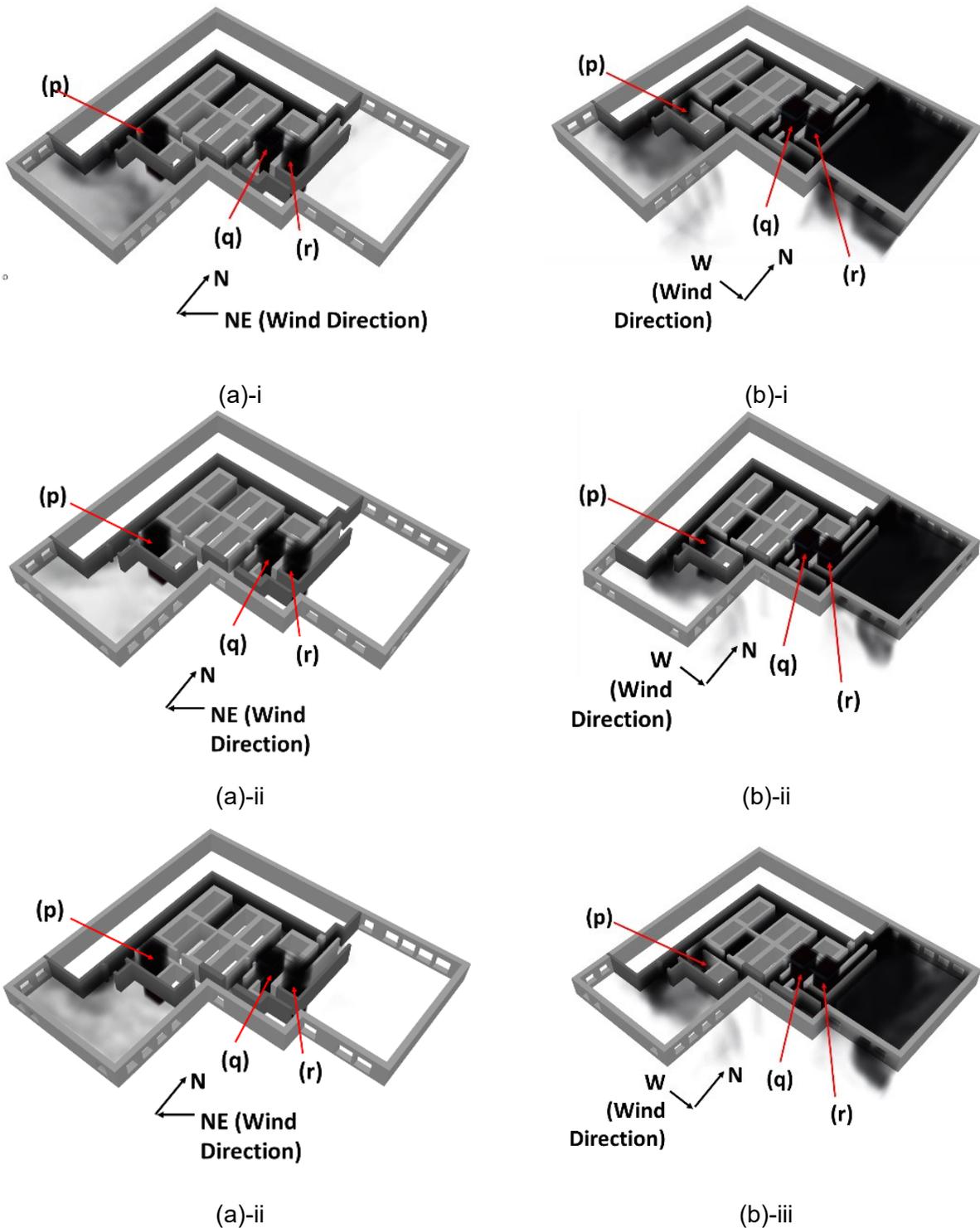

Figure 15: Smoke dispersion on refuge floor after 600 seconds of fire; (a)- wind is from North-East (b)- wind is from West; i-current setup, ii-1st modification, iii-2nd modification; p, q, r are the fire locations

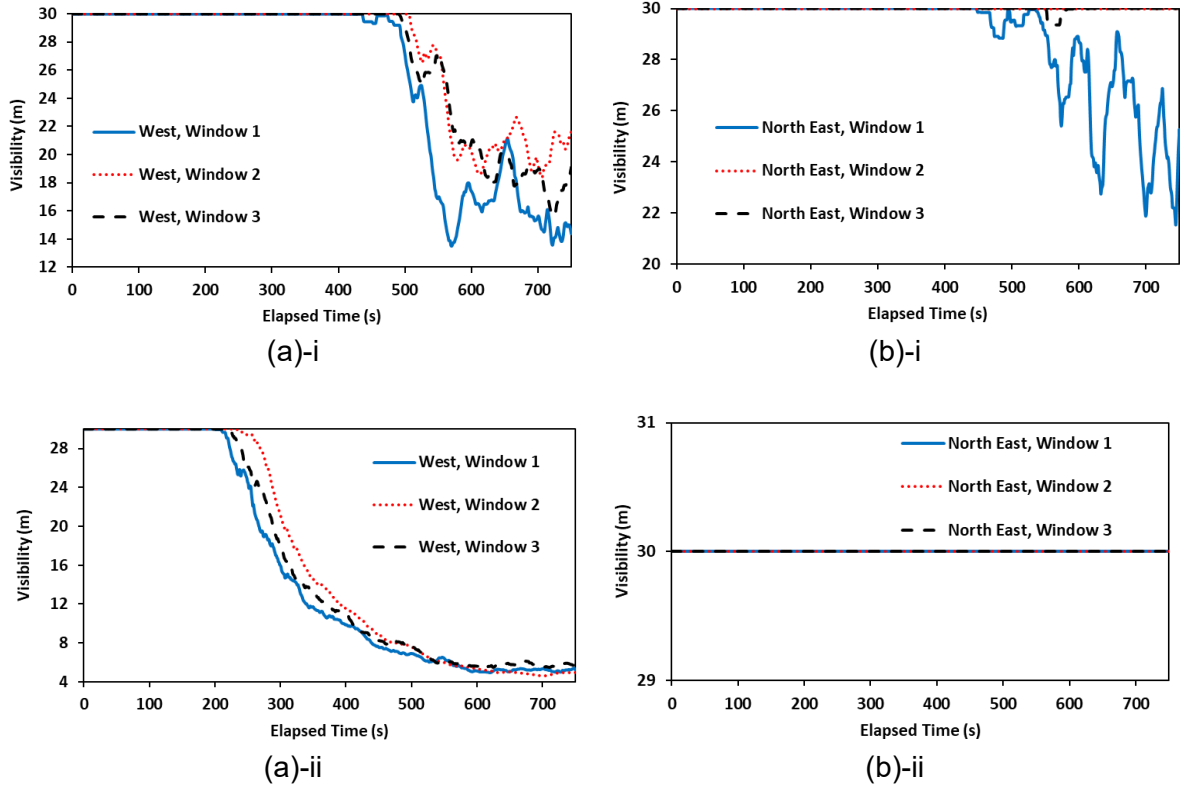

Figure 16: Variation of visibility with time of different window setups at 2m height; (a) readings from P2 in refuge area 1 (b) readings from P3 in refuge area 2; i-wind is from West, ii- wind is from North-East

**Velocity distribution within the refuge floor**

Behaviour of the wind inside the refuge floor and around the building was analysed using velocity vector profiles shown in figure 17. It is clear that wind flow was affected by the shape of the building and a void was created at the leeward side. The wind was attracted to the void when it was near that. Velocity distribution is shown in figure 17 (a) when the wind is from North-East. Air circulation inside refuge floor area 2 could be seen in configuration - 1 due to the windows in wall 06. Wind flow from wall 05 to wall 07 was higher in configuration - 2 because of the increased area of openings and the absence of windows in wall 06. Both modifications had higher wind velocity inside the refuge floor than the designed window setup in refuge area 1. Generated cross ventilation is a key attribute in terms of removing accumulated smoke.

Figure 17 (b) illustrates results for the West wind direction. Windows in wall 01 were introduced in the modifications to minimize the unbearable wind velocity in refuge floor area 1. And the results show that the solution proposed in the modifications is effective. Air was circulating inside refuge area 2 since windows failed to induce an airflow inside the floor. This could lead to severe smoke accumulation in refuge area 2. Modifications were applied to create airflow by introducing windows to wall 06 and increasing the window area of the windows in wall 05 and wall 07. Although there were some improvements, it was not enough to extract smoke from the refuge floor. It is clear that the orientation of this building must be changed to obtain benefits from natural ventilation. Otherwise, this building must use mechanical ventilation to create an airflow inside refuge area 2.

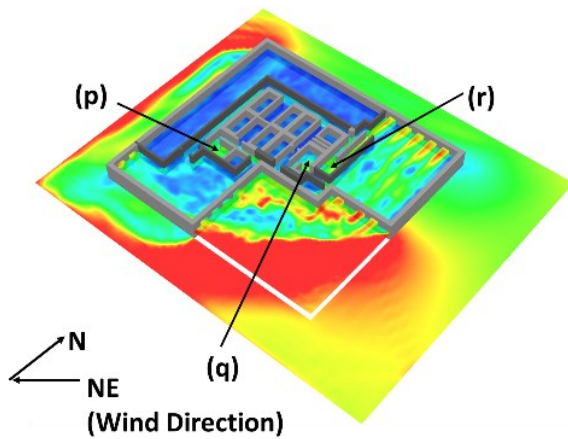
(a)-i

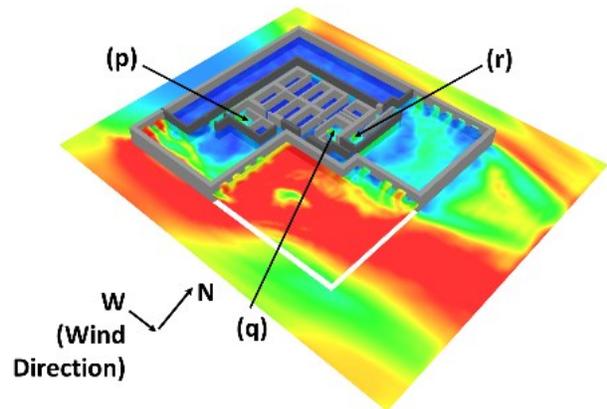
(b)-i

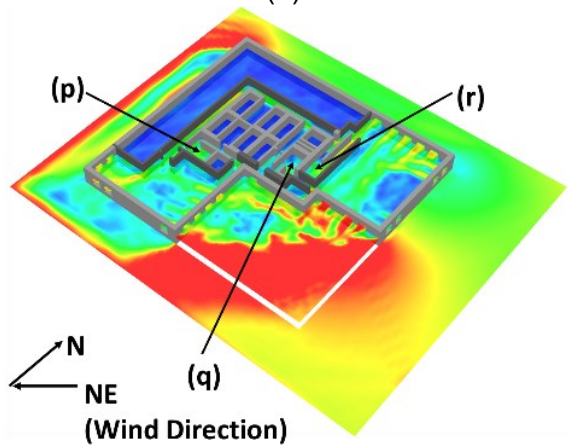
(a)-ii

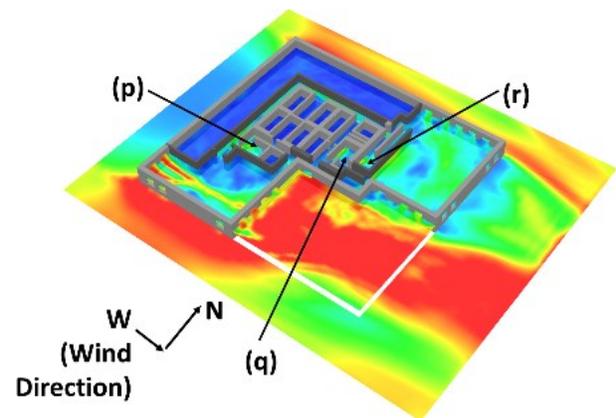
(b)-ii

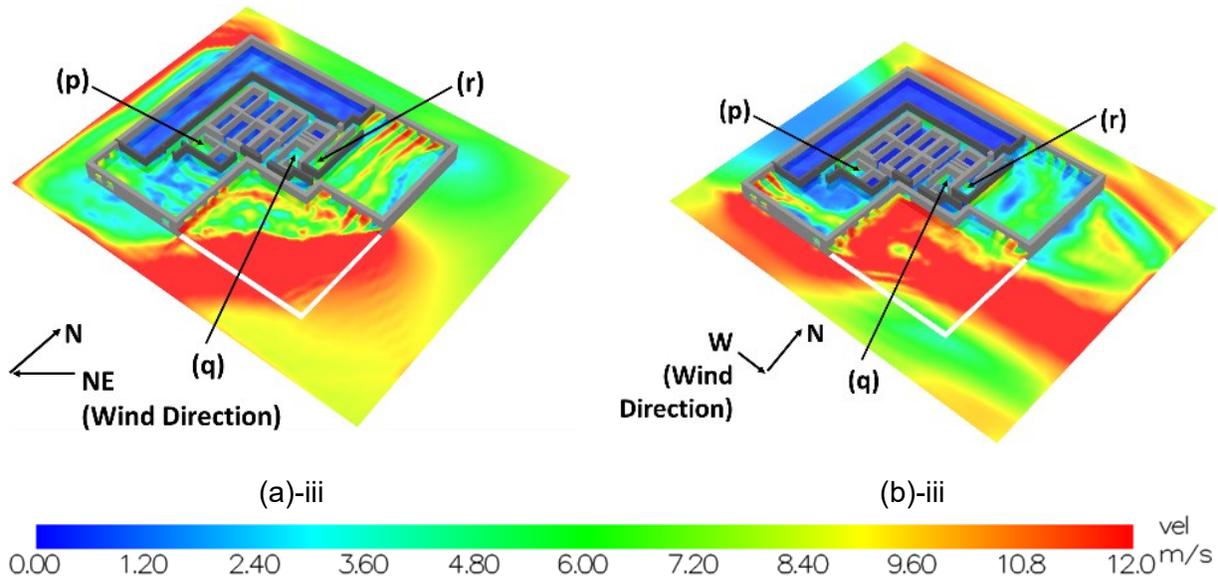

Figure 17: Velocity vector distribution on the refuge floor at 2 m height after 600 seconds of fire(a)- wind is from North-East (b)- wind is from West; i-current setup, ii-1st modification, iii-2nd modification; p, q, r are the fire locations

**Conclusion**

This paper represented the application of CFD in measuring habitable conditions inside the refuge floor of a high-rise building in case of an emergency situation. Mathematical models such as turbulence and combustion were validated to ensure the correct models were used for the simulation of case building. The validation was conducted for a 10-storey building and the results proved that the mathematical model was able to model an actual scenario. The intention of this research was to identify the ability of natural ventilation to remove any amount of smoke that could log inside the refuge floor. Therefore, the model was implemented to achieve the maximum amount of smoke on the refuge floor. Temperature CO concentration and visibility were selected as the parameters to discuss the survivability of the refuge floor. Apart from the designed window configuration, two modifications were suggested for the assessment. There could be seen a nearly stagnant flow inside the refuge floor area 2 when the wind blew from the West direction. The openings have failed to generate cross-ventilation since the windward side is blocked. This revealed the fact that the orientation of the window openings and wind direction should be taken into account when the ventilation strategy is designed. The results of refuge area 1 when the wind was moving from the West revealed the effectiveness of cross-ventilation to egress the smoke inside the refuge floor. The window arrangement of each configuration was capable of providing cross-ventilation when the wind blew from the North-East direction. Overall, considering the profiles of temperature, CO concentration, visibility, and smoke behaviour, configuration - 1 was better than configuration - 2 and designed setup.

Fire propagation through the service core was analysed in this study. Fire propagation from the outside of the building should be analysed. Also, the propagation of fire through different types of facades, analysis of insulation materials of the façade system, and the impact of the drencher system as a fire provision could be suggested for further research.

## Acknowledgment

Authors appreciate the support of Prof. R.U. Halwathura, Eng. Shiromal Fernando, Eng. U. Virajini, Mr. H.W. Kularathna, Eng. S. Amarasekara, Mr. E.M. Jayathilaka Banda, Mr. Panduka Wijeyawardena, and Mr. Lakmal Kandanage for the research. Also, authors acknowledge the financial support provided by University of Moratuwa.